# A hybrid molecular dynamics/atomic-scale finite element method for quasi-static atomistic simulations at finite temperature

Ran Xu, Bin Liu[*]

*AML, CNMM, Department of Engineering Mechanics, Tsinghua University, Beijing 100084, China*

* Corresponding author. Tel.: 86-10-62786194.

E-mail address: liubin@tsinghua.edu.cn (B. Liu).

## ABASTRACT

In this paper, a hybrid quasi-static atomistic simulation method at finite temperature is developed, which combines the advantages of MD for thermal equilibrium and atomic-scale finite element method (AFEM) for efficient equilibration. Some temperature effects are embedded in static AFEM simulation by applying the virtual and equivalent thermal disturbance forces extracted from MD. Alternatively performing MD and AFEM can quickly obtain a series of thermodynamic equilibrium configurations such that a quasi-static process is modeled. Moreover, a stirring-accelerated MD/AFEM fast relaxation approach is proposed, in which the atomic forces and velocities are randomly exchanged to artificially accelerate the "slow processes" such as mechanical wave propagation and thermal diffusion. The efficiency of the proposed methods is demonstrated by numerical examples on single wall carbon nanotubes.

***KEYWORDS***: Quasi-static; Finite temperature; Molecular dynamics; Atomic-scale finite element method.

## 1. Introduction

Materials with morphological features in nanoscale like nanoparticles, nanotubes and nanoshells possess many special properties and have potential applications in microelectromechanical systems and composites (Sanchez-Portal et al., 1999; Demczyk et al., 2002; Zheng and Jiang, 2002; Boland et al., 2005; Huang et al., 2006; Ma et al., 2009). These properties are influenced by many factors, such as surface/interface, size, temperature etc., which pose a huge challenge for theoretical models to accurately account for all effects. Moreover, experimental study on materials with such small scale is not an easy task since the measurement might have large error due to comparable environment noise. Alternatively, many numerical methods have been dedicated to performance prediction of nanomaterials, and the molecular dynamics (MD) method (Alder and Wainwright, 1957, 1959) is the most direct and widely-used numerical approach. To avoid the cumulative errors of explicit numerical integration, the time step of MD must be at least one order smaller than the atomic vibration period, i.e. 1 femtosecond ($10^{-15} s$) for realistic molecular systems. Due to such small time step, the huge amount of computations limits the temporal scale of MD simulations, and therefore many physical/chemical processes have to be virtually accelerated in MD. For example, the strain rates used in MD simulations of material deformation are unrealistically higher, in the order of $10^7 s^{-1}$ (Branício and Rino, 2000; Wei et al., 2003; Koh et al., 2005; Chen et al., 2011; Jennings et al., 2011), in contrast to the most experimental strain rates less than $10^{-4} s^{-1}$ recommended by standard mechanical test procedure, in which materials always undergo the approximately or quasi- static external loadings. Even in split Hopkinson pressure bar experiments on nanomaterials-based composites test (Al-Lafi et al., 2010; Lim et al., 2011), the highest strain rate is only in the order of $10^4 s^{-1}$, still much lower than the rate in MD. Therefore, this huge difference on the strain rate always casts a doubt on the validity of MD simulation results. Many researchers have been working over the past decades to extend the computational limit of MD in temporal scale (E et al., 2003; Iannuzzi et al., 2003; Dupuy et al., 2005; Zhou, 2005; Kulkarni et al., 2008; Branduardi et al., 2012). The heterogeneous multi-scale method proposed by E et al. (2003), is a both spatial- and temporal- multi- scale method for equilibrium simulation, in which the macro-scale variables are achieved from the microscopic simulation of local limited spatial/temporal domains.

Actually, it is noted that most mechanical tests can be viewed as quasi-static processes, which can be simulated by the molecular mechanics (MM) method. For a specific loading value, MM essentially performs the potential energy minimization, and obtains the corresponding static response of molecular systems. Conjugate gradient method and steepest descent method are two traditional algorithms used in MM simulations, but their efficiencies are lower and on order- $N^2$. Liu et al. (2004, 2005) developed the atomic-scale finite element method (AFEM), which is an order-$N$

MM method for discrete atoms with high efficiency. However, an obvious drawback of all MM methods is that they do not involve temperature and can only accurately simulate the atomic system at $0K$. Many researchers tried to introduce the effect of temperature into MM theory to build a modified energy minimization at finite temperature. Lesar et al. (1989) presented a finite-temperature atomic structure simulation technique by minimizing the free energy of solid. Najafabadi and Srolovitz (1995) studied the accuracy of local harmonic approximation with different interatomic potential functions. Wang et al. (2008) proposed a systematic molecular/cluster statistical thermodynamics method based on Helmholtz free energy. Furthermore Jiang et al. (2005) combined local harmonic approximation and atomistic-based continuum theory, and studied graphene and diamond at different temperature. The variational coarse-graining approach proposed by Kulkarni et al. (2008), can simulate thermo-dynamic equilibrium properties of atomic system at finite temperature, in which the approximate probability distribution and partition functions of atoms are derived from variational mean-field theory and maximum-entropy method, and the classical empirical atomic interaction potential functions and temperature are taken as inputs. But for the nonlinear and nonlocal force field, these statistical methods may deviate from realistic state under higher temperature due to the harmonic oscillation approximation. Iacobellis and Behdinan (2012) proposed a spatial multi-scale static simulation method combining AFEM and continuum finite element method (FEM), and the temperature-dependent potential functions of atomic interaction are implemented, in which the temperature effects are considered by the coefficient of thermal expansion. Guo and Chang (2002) developed a freezing atom method (FAM) to study the quasi-static responses of molecular systems by alternatively performing MM and MD methods. In their numerical experiments, FAM is proved more efficient than classical MD, but the temperature effects, such as thermal expansion, are completely ignored in its MM simulation stage, which might induce jump at the transition of algorithms and increase the computation time.

The purpose of this paper is to establish a quasi-static (i.e. the strain rate is very close to zero) finite temperature molecular simulation method directly from interatomic potential functions without any approximated statistical treatments. In particular, it is a hybrid molecular dynamics/atomic-scale finite element method. MD simulation method is performed to ensure the thermal equilibrium of an atomic system, and AFEM accounting for the temperature effects is performed to accelerate the convergence of its equilibration. The paper is organized as follows. Section 2 provides a brief introduction of molecular dynamics method and atomic-scale finite element method. For an initially thermodynamic equilibrium molecular system, Section 3 presents a detailed hybrid MD/AFEM atomistic simulation method to model quasi-static deformation process at finite temperature. For an initially non-equilibrium molecular system, a stirring-accelerated MD/AFEM relaxation method for quick balance at finite temperature is proposed in section 4. The conclusions are summarized in Section 5.

## 2. Brief introductions to molecular dynamics method and atomic-scale finite element method

The atoms in molecular simulations are described as virtual spatial points with the associated mass, and the interactions among atoms are usually described by empirical potential functions. The basic potential functions include both the bonded part, i.e., the interaction between bonded atoms, and the non-bonded part such as the long-distance Van der Waals force and Coulomb force. A potential energy of molecular system can be expressed as a function of the positions of all ( $N$ ) atoms,

$$E_{\text{total}} = U_{\text{total}}\left(\boldsymbol{x}_1, \boldsymbol{x}_2, ..., \boldsymbol{x}_N\right) - \sum_{i=1}^{N} \boldsymbol{f}_i^{\text{ext}} \cdot \boldsymbol{x}_i, \qquad (1)$$

where $\boldsymbol{x}_i$ is the position of atom $i$, $U_{\text{total}}$ is the sum of atomic interaction energy, and $\boldsymbol{f}_i^{\text{ext}}$ is the external force (if there is any) exerted on atom $i$. The non-balanced force $\boldsymbol{f}_i$ of atom $i$ is

$$\boldsymbol{f}_i = -\frac{\partial E_{\text{total}}}{\partial \boldsymbol{x}_i} = -\frac{\partial U_{\text{total}}}{\partial \boldsymbol{x}_i} + \boldsymbol{f}_i^{\text{ext}}. \qquad (2)$$

The theoretical basis of molecular dynamics method is principally the following second Newton's law of motion,

$$\frac{d^2 \boldsymbol{x}_i}{dt^2} = \frac{\boldsymbol{f}_i}{m_i}, \qquad (3)$$

where $m_i$ is the associated mass of atom $i$. In MD simulations, equation (3) is solved by explicit numerical methods, such as Verlet integration algorithm (Verlet, 1967, 1968)

$$x_i(t+\Delta t) = 2x_i(t) - x_i(t-\Delta t) + \frac{f_i(t)}{m_i}\Delta t^2$$
$$\dot{x}_i(t+\Delta t) = \frac{x_i(t+\Delta t) - x_i(t-\Delta t)}{2\Delta t} \quad , \qquad (4)$$

and leap-frog integration algorithm (Hockney and Eastwood, 1988)

$$\dot{x}_i(t+\Delta t/2) = \dot{x}_i(t-\Delta t/2) + \frac{f_i(t)}{m_i}\Delta t$$
$$x_i(t+\Delta t) = x_i(t) + \dot{x}_i(t+\Delta t/2)\Delta t \quad , \qquad (5)$$

where $\Delta t$ is the time step. It should be pointed out that the time step must be very small to ensure this explicit integration algorithm stable. Thus the amount of computation of MD simulation is very huge, which results in the temporal scale of MD simulation far smaller than those of actual processes.

The atomic-scale finite element method (AFEM) proposed by Liu et al. (2004, 2005) is an efficient MM simulation algorithm. The potential energy (Eq. (1)) can be rewritten as the second-order Taylor expansion around the initial guessed positions $\mathbf{X}^{(0)} = \left(x_1^{(0)}, x_2^{(0)}, ..., x_N^{(0)}\right)$ of atoms

$$E_{\text{total}} = E_{\text{total}}\left(\mathbf{X}^{(0)}\right) + \left.\frac{\partial E_{\text{total}}}{\partial \mathbf{X}}\right|_{\mathbf{X}=\mathbf{X}^{(0)}} \cdot \left(\mathbf{X}-\mathbf{X}^{(0)}\right)$$
$$+ \frac{1}{2}\left(\mathbf{X}-\mathbf{X}^{(0)}\right) \cdot \left.\frac{\partial^2 E_{\text{total}}}{\partial \mathbf{X}\partial \mathbf{X}}\right|_{\mathbf{X}=\mathbf{X}^{(0)}} \cdot \left(\mathbf{X}-\mathbf{X}^{(0)}\right) + O\left(\left(\mathbf{X}-\mathbf{X}^{(0)}\right)^3\right) \qquad (6)$$

The state of minimal potential energy corresponds to

$$\frac{\partial E_{\text{total}}}{\partial \mathbf{X}} = 0 \,. \qquad (7)$$

Substituting Eq. (6) into Eq. (7) yields the governing equation of AFEM as,

$$\mathbf{KU} = \mathbf{F} \,, \qquad (8)$$

where $\mathbf{U} = \left(\mathbf{X} - \mathbf{X}^{(0)}\right)$ is the displacement vector, $\mathbf{F} = \left(f_1^{(0)}, f_2^{(0)}, ..., f_N^{(0)}\right)$ is the non-equilibrium force vector at $\mathbf{X}^{(0)}$, and

$$\mathbf{K} = \left.\frac{\partial^2 E_{\text{total}}}{\partial \mathbf{X} \partial \mathbf{X}}\right|_{\mathbf{X} = \mathbf{X}^{(0)}} \qquad (9)$$

is the stiffness matrix. Equation (8) will be solved iteratively until $\mathbf{F}$ meets the specific accuracy requirement. It can be found that AFEM does not involve any approximations of conventional FEM (e.g., shape functions), and is as accurate as molecular mechanics simulations. Because AFEM uses both first and second derivatives of potential energy (Eq. (1)) in the minimization computation, previous numerical examples (Liu et al. 2004, 2005) have demonstrated that it is an order-$N$ method, much faster than the widely used order-$N^2$ conjugate gradient method, and is suitable for large scale computations.

## 3. A hybrid quasi-static method of atomistic simulation at finite temperature

### 3.1 The algorithm

As we know, using MM relaxation before MD simulation is a typical and widely used strategy to accelerate the convergence of MD. The underlying mechanism is quickly searching for a state close to thermodynamic equilibrium first, and then performing time consuming MD simulation. The basic idea of our algorithm is frequently using MM relaxation during MD simulation to model a quasi-static loading process efficiently. However, if traditional MM is adopted in subsequent static relaxation, the molecular system will be cooled down back to 0K, and the subsequent

MD simulation has to heat up the system again, as suggested by Guo and Chang (2002). Then the simulation process is still time consuming. Moreover, since temperature effects, such as thermal expansion, are included in MD but not included in traditional MM, the dramatic temperature-induced-change of the molecular system at the transition of MD and MM is not realistic, which might fail to reproduce a quasi-static loading process at a constant temperature. To overcome this drawback, we first develop a revised MM method (AFEM is adopted in this paper) capable of accounting for temperature effects.

In molecular systems at finite temperature, atoms usually deviate from their static equilibrium positions due to thermal motion. This temperature-induced deviation can be equivalently viewed as the results of thermal disturbance forces, which represent the forces needed to keep the atoms still and equilibrium in the current positions when the temperature or velocity is removed. The thermal disturbance forces change over time, and it is difficult to determine them in a theoretical way.

In the following, we propose a method to extract the thermal disturbance forces from MD simulations directly. For a thermodynamic equilibrium system shown in Fig.1a, the opposition of the net force of atom $i$ obtained by Eq. (2), $-f_i$, should be applied to atom $i$ to freeze the current positions of atoms, and obviously an equivalent static equilibrium can be reached for the configuration at this moment (see Fig. 1b). Therefore, $\mathbf{F}^{\text{thermal}} = \left( f_1^{\text{thermal}}, f_2^{\text{thermal}}, ..., f_N^{\text{thermal}} \right) = \left( -f_1^{(0)}, -f_2^{(0)}, ..., -f_N^{(0)} \right)$ are the thermal disturbance forces, and will be applied to the molecular system as additional loadings in AFEM static simulation to reflect the temperature effects.

Correspondingly, the MD simulation transfers to AFEM static simulation smoothly. When the external forces are changed, a new quasi- thermodynamic equilibrium configuration can be quickly obtained via AFEM. We then "melt" the system and perform a few steps of MD relaxation. The detailed scheme is schematically shown in Fig. 1 and given as follows.

(1). *Freezing (from Fig.1a to b)*. We first freeze the MD simulation process of a thermodynamic equilibrium molecular system, and store/compute the current positions $\mathbf{X}^{(0)} = \left( \boldsymbol{x}_1^{(0)}, \boldsymbol{x}_2^{(0)}, ..., \boldsymbol{x}_N^{(0)} \right)$, velocities $\dot{\mathbf{X}}^{(0)} = \left( \boldsymbol{v}_1^{(0)}, \boldsymbol{v}_2^{(0)}, ..., \boldsymbol{v}_N^{(0)} \right)$ of atoms, and the current thermal disturbance forces $\mathbf{F}^{\text{thermal}} = \left( -\boldsymbol{f}_1^{(0)}, -\boldsymbol{f}_2^{(0)}, \cdots, -\boldsymbol{f}_N^{(0)} \right)$.

(2). *Loading via AFEM (from Fig.1b to c)*. When the external force has a increment $\Delta \mathbf{F}^{\text{ext}}$, the real non-equilibrium forces of the atoms become $\mathbf{F}^{\text{new}}$, together with the virtual and equivalent thermal disturbance forces $\mathbf{F}^{\text{thermal}}$, the governing equations of AFEM with temperature effects is

$$\mathbf{KU} = \mathbf{F}^{\text{new}} + \mathbf{F}^{\text{thermal}}, \tag{10}$$

Equation (10) will be solved iteratively, and the positions of atoms are updated by $\mathbf{X}^{(k+1)} = \mathbf{X}^{(k)} + \mathbf{U}$ until the accuracy requirement is met.

(3). *Melting (from Fig.1c to d)*. Give the stored velocity $\dot{\mathbf{X}}^{(0)}$ back to each atom and remove the thermal disturbance forces.

(4). *Thermodynamic relaxation via MD (from Fig.1d to e)*. Proceed with the MD simulation under the desired loading and temperature until reaching a new thermodynamic equilibrium state.

Through alternatively performing the above four-step hybrid MD/AFEM

algorithm, we can quickly model a series of thermodynamic equilibrium states under different loading, i.e., a quasi-static atomistic simulation is realized.

It should be emphasized that if the external forces remain unchanged, the AFEM simulation with temperature effects in Step 2 will yield a zero displacement vector since $\mathbf{F}^{new} + \mathbf{F}^{thermal} = 0$ in Eq. (10). Therefore, the inserting AFEM simulation in this case does not have any influence on MD simulation, which implies complete smoothness between MD and AFEM simulation. If the external forces are changed, AFEM relaxation with temperature effects will yield a new system state very close to thermodynamic equilibrium, and only small amount of subsequent MD relaxation is needed for final equilibrium. This hybrid MD/AFEM algorithm is then much faster than traditional MD.

Another thing should be mentioned is the accuracy requirement of AFEM simulation in Step 2. The thermal disturbance forces for the atoms in a thermodynamic equilibrium system exhibits some random features in magnitude and orientation, and their resultant force and moment should be close to zero. After several iteration of AFEM simulation, the resultant force $\mathbf{P} = \sum \boldsymbol{f}_i^{(0)}$ is still close to zero, but the resultant moment $\mathbf{M} = \sum \boldsymbol{x}_i^{(k)} \times \left( \boldsymbol{f}_i^{(0)} \right)$ might deviate from zero since the atom positions are changed. Therefore, we recommend that only a few iterations (less than 5) of AFEM relaxation are adopted.

In many simulations, the validity of simulation parameters need test. For example, in finite element method, the mesh density is an important simulation parameter. If several different mesh densities yield the converged result, it means the mesh density is proper. Similarly, the loading increment and MD relaxation period are the simulation parameters in the proposed method, and we may run several simulations with different parameters to test their validity. Besides this accelerating relaxation treatment, the proposed hybrid AFEM/MD method has similar problems as the traditional MD does. No one knows how long MD relaxation should be used in reality to avoid missing rare events. We suggest that trying several different MD relaxation periods and then selecting acceptable one from both accuracy and efficiency aspects might be a compromise option.

## 3.2 Numerical experiments

Carbon nanotube (CNT) has extraordinary mechanical properties, and has attracted much research interest. Some researchers predicted that the stiffness and strength of CNT are temperature dependent (Yakobson et al., 2006; Wang et al., 2009). But, the predicted tensile strength (failure strain) is far above experimental value due to unrealistically high strain rate in MD simulation (Yakobson et al., 1997; Ruoff et al., 2000; Jose-Yacaman et al., 2003). In the following, we use the proposed hybrid MD/AFEM algorithm to simulate quasi-static behavior of (8,8) single wall CNT under axial tension at different temperature. The hybrid AFEM/MD computation is executed with a serial code developed with Intel® Visual FORTRAN Compiler. In the AFEM relating code, the PARDISO package in Intel® MKL is employed to solve Eq. (10) in Step 2.

The widely-used second-generation Tersoff-Brenner multi-body potential function is adopted to model Carbon-Carbon (C-C) interactions (Brenner et al., 2002). Initially, the CNT has reached thermodynamic equilibrium state at a specific temperature. The increasing tensile load is then imposed to CNT step-by-step by the proposed quasi-static simulation approach: the end atoms move $0.05\text{nm}$ along axial direction in each AFEM simulation stage, and the following $20\text{ps}$ MD relaxation is performed to reach the thermodynamic equilibrium state with Nose-Hoover temperature coupling (Nose, 1984; Hoover, 1985). The potential energy of this (8,8) CNT under increasing tensile loading at $T = 150\text{K}$ and $T = 300\text{K}$ is shown in Fig. 2. The data of $-20\text{ps} \sim 0\text{ps}$ shows the fluctuations of potential energy at the load-free thermodynamic equilibrium state, which can be viewed as a reference state. It is found that the fluctuation of potential energy in all MD relaxation is lower than 5% of the mean value of potential energy, and is similar to that of the reference state, which implies thermodynamic equilibrium. Moreover, the results in Fig. 2 indicate that the AFEM used in loading step can directly find a good dynamic configuration, and the subsequent MD relaxation simulation can rapidly reach equilibrium state with very tiny computational cost. Therefore, this numerical example demonstrates that the proposed hybrid MD/AFEM algorithm can realize the quasi-static simulation at finite temperature.

Figure 3a shows the temperature dependence of the tensile stress-strain curves of (8,8) CNT obtained from the proposed quasi-static simulation method. In order to avoid the controversy on the thickness of CNT, we choose the following 2D stress definition $\sigma^{2D}$ proposed by Hone et al. (2008)

$$\sigma^{2D} = F^{Ext}/2\pi r, \qquad\qquad (11)$$

where $F^{Ext}$ is the total axial tensile force imposed on CNT, and $r$ is the nominal radius of CNT. Obviously, the hybrid MD/AFEM simulation can reasonably capture the temperature effects on the tensile behaviors, and predict that the tensile elastic modulus and the strength of the CNT decrease as the increasing of temperature (see Fig.3a, b).

The efficiency of this hybrid quasi-static method on single core/CPU has been tested by simulating the CNTs with different number of atoms. The initially straight CNTs are first heated to 300K, and are then subject to the same lateral force 3000 kJmol$^{-1}$nm$^{-1}$ in the middle with two ends fixed. Nose-Hoover thermostat is used in MD with reference temperature $T_0 = 300$K and the period $\tau_T = 0.1$ps. The convergence criterion is chosen as the fluctuations of the deflection in CNTs less than a threshold value. Figure 4 shows the required CPU time of the hybrid AFEM/MD method and the pure MD in simulating (5,5) armchair CNTs with 400, 800, 1600, 3200, 6400, 12800, 25600 and 51200 atoms. The CPU time of the hybrid AFEM/MD method, denoted by open circles, displays an approximately linear dependence on number of atoms, very close to the linear fitting curve, $Time = 0.1328N$. It is known that AFEM used in Step 2 is an order-$N$ method (Liu et al., 2005). The MD relaxation after AFEM loading process is only used to tackle the local non-equilibrium, and the required number of MD integration steps is independent with the number of atoms. Thus, the hybrid quasi-static method is an order-$N$ method, and therefore has high efficiency. However, the overall time consumption of MD, denoted by the open boxes in Fig.4, is much higher than order-N (with approximately quadratic fitting relation $\text{Time} = 0.0009581N^2$). That is because in pure MD, larger systems need longer physical relaxation time, and in each time step the time consumption is on order-N.

When the number of atoms is large, MD can be parallelized to use thousands CPU/cores to simulate large or complex molecular systems. We have developed a hierarchical parallel algorithm for the solution of super large-scale sparse linear equations (Xu R et al., 2013). It has good parallel efficiency and can deal with more than one billion unknowns in implicit FEM. Therefore, the hybrid AFEM/MD method can also be implemented on clusters.

## 4. A stirring-accelerated MD/AFEM fast relaxation method

Section 3 presents an algorithm for quasi-static atomistic simulation at finite temperature, which is essentially a series of processes starting from an old thermodynamic equilibrium state and searching for a new equilibrium state. However, obtaining the initial thermodynamic equilibrium state of molecular systems sometimes needs a lot of computation effort. Usually, mechanical motion (e.g., vibration and wave propagation) and thermal motion exist in most dynamic systems, and their natural decay may take a very long time, which is governed by some intrinsic "slow process" mechanisms, such as wave velocity and dissipation rate. To artificially accelerate this slow relaxation process, as demonstrated in Section 3, if a set of reasonable thermal disturbance forces can be obtained and introduced into hybrid MD/AFEM simulation, the thermodynamic equilibrium state will be quickly achieved. However, the disturbance forces extracted from non-equilibrium molecular systems can not be directly used, since they have some strong correlation due to the mechanical motion of local atom clusters. Noting that the thermal disturbance forces are more random with less correlation, we propose converting the original disturbance forces into quasi-thermal disturbance-forces via randomly "stirring". In particular, the disturbance forces and velocities are randomly exchanged among atoms, and this treatment is equivalent to making the velocities of wave propagation and thermal diffusion infinite. Therefore, the relaxation process can be significantly shortened. Actually, our "stirring" method draws inspiration from previous works, such as Andersen (1980). In his thermal bath algorithm, the atomic velocities are randomly assigned.

## 4.1 The basic algorithm

Figure 5 shows the schematics of the "stirring" approach, which consists of the following four looped steps:

(1). *Freezing (from Fig.5a to b)*. Store the positions $\mathbf{X}^{(0)} = \left( \boldsymbol{x}_1^{(0)}, \boldsymbol{x}_2^{(0)}, ..., \boldsymbol{x}_N^{(0)} \right)$ and velocities $\dot{\mathbf{X}}^{(0)} = \left( \boldsymbol{v}_1^{(0)}, \boldsymbol{v}_2^{(0)}, ..., \boldsymbol{v}_N^{(0)} \right)$, and calculate the disturbance forces $\mathbf{F}^{\text{disturbance}} = \left( \boldsymbol{f}_1^{\text{disturbance}}, \boldsymbol{f}_2^{\text{disturbance}}, ..., \boldsymbol{f}_N^{\text{disturbance}} \right) = \left( -\boldsymbol{f}_1^{(0)}, -\boldsymbol{f}_2^{(0)}, ..., -\boldsymbol{f}_N^{(0)} \right)$ caused by the thermal oscillation and mechanical motion.

(2). *Stirring of disturbance forces and AFEM relaxation (from Fig.5b to c)*. The disturbance forces $\mathbf{F}^{\text{disturbance}}$ will be revised by "stirring" as,

$$\mathbf{F}^{\text{stired}} = \underset{i=1}{\overset{3N}{\text{random}}} \left( \mathbf{F}^{\text{disturbance}} \right), \qquad (12)$$

where random$(x)$ is a random assigning function, in which the $3N$ ($N$ is the number of atoms) components of vector $\mathbf{F}^{\text{disturbance}}$ are randomly rearranged. Then the AFEM simulation can be performed by iteratively solving the following equation,

$$\mathbf{KU} = \mathbf{F}^{\text{new}} + \mathbf{F}^{\text{stired}}. \qquad (13)$$

(3). *Stirring of velocities and melting (from Fig.5c to d).* Similarly, the velocities of atoms are stirred to eliminate the correlation of non-equilibrium thermal and mechanical motions, i.e.,

$$\dot{\mathbf{X}}^{\text{stired}} = \underset{i=1}{\overset{3N}{\text{random}}} \left( \dot{\mathbf{X}}^{(0)} \right). \qquad (14)$$

The stirred disturbance forces are removed, and the stirred velocities are then assigned to atoms. The system is melted into a dynamical one.

(4). *Thermodynamic MD relaxation (from Fig.5d to e).* A few of MD integration steps ($\leq 200$) is done under temperature coupling for relaxation.

The molecular system will reach a thermodynamic equilibrium state after a few of loops.

## 4.2 Numerical experiments

We still simulate mechanical behaviors of CNTs to demonstrate the efficiency of the proposed stirring-accelerated MD/AFEM relaxation method. As shown in Fig. 6a, a axial displacement loading is initially applied on the middle of an (8,8) CNT with fixed ends, and the system is in thermodynamic equilibrium state at $T = 300\text{K}$. If the loading is suddenly released, the CNT will start to vibrate. The CNT oscillator is first simulated as an isolated system in pure MD simulation, and the gradual decrease of the potential energy fluctuation versus time is shown in Fig. 6b. It is found that the quality factor $Q$ of this tensile-compressive oscillator is $1.1105 \times 10^4$ based on the definition from Jiang et al. (2004), which is an extraordinarily high quality factor implying very slow decay of the mechanical vibration. Therefore, it will take a long time in MD simulation to reach the final equilibrium state.

For comparison, Fig. 6b, c, and d show the variation of potential energy of the CNT oscillator from $0\text{ps}$ to $20\text{ps}$ of three relaxation simulations (the displacement loading is released at 0ps): MD without any interference (i.e., free vibration), MD with Nose-Hoover thermostat, and stirring-accelerated MD/AFEM relaxation. The dot lines in each figure are the average potential energy of equilibrium non-deformed CNT at $300\text{K}$. The fluctuation of potential energy before 0ps shown in each figure is used as a reference. In the free vibration simulation (see Fig. 6b), the potential energy fluctuation slowly decays as time increasing, and the vibration frequency is nearly the same as its intrinsic vibration frequency. The Nose-Hoover thermostat can reduce the fluctuations of potential energy to some extent (see Fig. 6c), but the amplitude after 20ps relaxation is still obviously larger than the reference fluctuation in -2.5ps to 0ps, which indicates that the thermodynamic equilibrium has not been reached. The reason is because that the thermostat only adjusts the temperature by rescaling the atomic velocities, and can not eliminate the mechanical vibration efficiently. In contrast, the stirring-accelerated MD/AFEM relaxation method can suppress the mechanical vibration quickly and the fluctuation of potential energy becomes on the same order of the reference equilibrium state as shown in Fig. 6d. In this simulation, the "stirring" is just performed every 0.02ps within the first 0.2ps, and the following simulation is only traditional MD with Nose-Hoover thermostat. It can be observed that less than 0.3ps, the mechanical vibration has been completely eliminated and the new thermodynamic equilibrium state has been achieved. We compare the efficiencies of the proposed hybrid method and pure MD with Nose-Hoover thermostat. It is found that for this small system, MD/AFEM is two orders faster than pure MD (12.67s VS. 670.72s). For larger systems, pure MD needs longer relaxation time, and the efficiency of MD/AFEM will become more prominent.

## 5. Disscussions and Conclusions

Theoretically, MD can do all molecular simulations. But in some cases, such as a material system under quasi-static varying loading, MD simulation can not be finished in a tolerable period, especially for relatively larger systems. A hybrid quasi-static atomistic simulation method at finite temperature is developed in this paper, which combines the advantages of MD for thermal equilibrium and AFEM for efficient equilibration. The thermal disturbance forces are extracted from MD and applied in AFEM to smoothly pass the temperature effect. Furthermore, a stirring-accelerated MD/AFEM relaxation approach is proposed to quickly obtain a thermodynamic equilibrium state by artificially accelerating the "slow relaxation processes" such as mechanical wave propagation and thermal diffusion. In this sense, the importance of our hybrid AFEM/MD method is providing a "fast forward button" besides a regular play mode (traditional MD). When an atomics system is under a quasi-static loading, the most part of time in this almost infinitely long process is relaxation. Therefore, via accelerating relaxation, the proposed hybrid AFEM/MD method can increase the efficiency dramatically, which has been demonstrated by the simulations on mechanical behaviors of carbon nanotubes. The proposed method therefore can make a quasi-static molecular simulation possible. We believe that the proposed quasi-static atomistic simulation method with temperature effect will play an important role in studying material behaviors.

**Acknowledgements**

The authors are grateful for the support from National Natural Science Foundation of China (Grant Nos. 90816006, 11090334, 10820101048, and 51232004), National Basic Research Program of China (973 Program) Grant No. 2010CB832701, Tsinghua University Initiative Scientific Research Program (No. 2011Z02173), and China Postdoctoral Science Foundation funded project (No. 2013M530606).

**Figure captions**

Figure 1 Schematic flow chart of hybrid molecular dynamics/atomic-scale finite element method for quasi-static atomistic simulations at finite temperature.

Figure 2 The potential energy variation of (8,8) CNT in step-by-step tensile test simulated by the hybrid MD/AFEM atomic quasi-static algorithm at $T$=150K and $T$=300K.

Figure 3 (a) The static 2D axial tensile stress as function of tensile strain for (8,8) CNT at different temperatures; (b) The temperature dependence of the tensile modulus of (8,8) single-wall CNT.

Figure 4 The required CPU time of the hybrid AFEM/MD method and the pure MD to acquire the steady deformed configurations of initially straight (5,5) armchair carbon nanotubes subject to the lateral force.

Figure 5 Schematic flow chart of stirring-accelerated MD/AFEM relaxation method for non-equilibrium molecular systems.

Figure 6 (a) Schematic diagram of an (8,8) single wall CNT subject to a axial displacement loading in the middle, and the displacement loading is suddenly released at 0ps. The variation of potential energy of CNT versus time of three relaxation simulations: MD without any interference (i.e., free vibration) (b), MD with Nose-Hoover thermostat at 300K (c), and stirring-accelerated MD/AFEM relaxation (d).

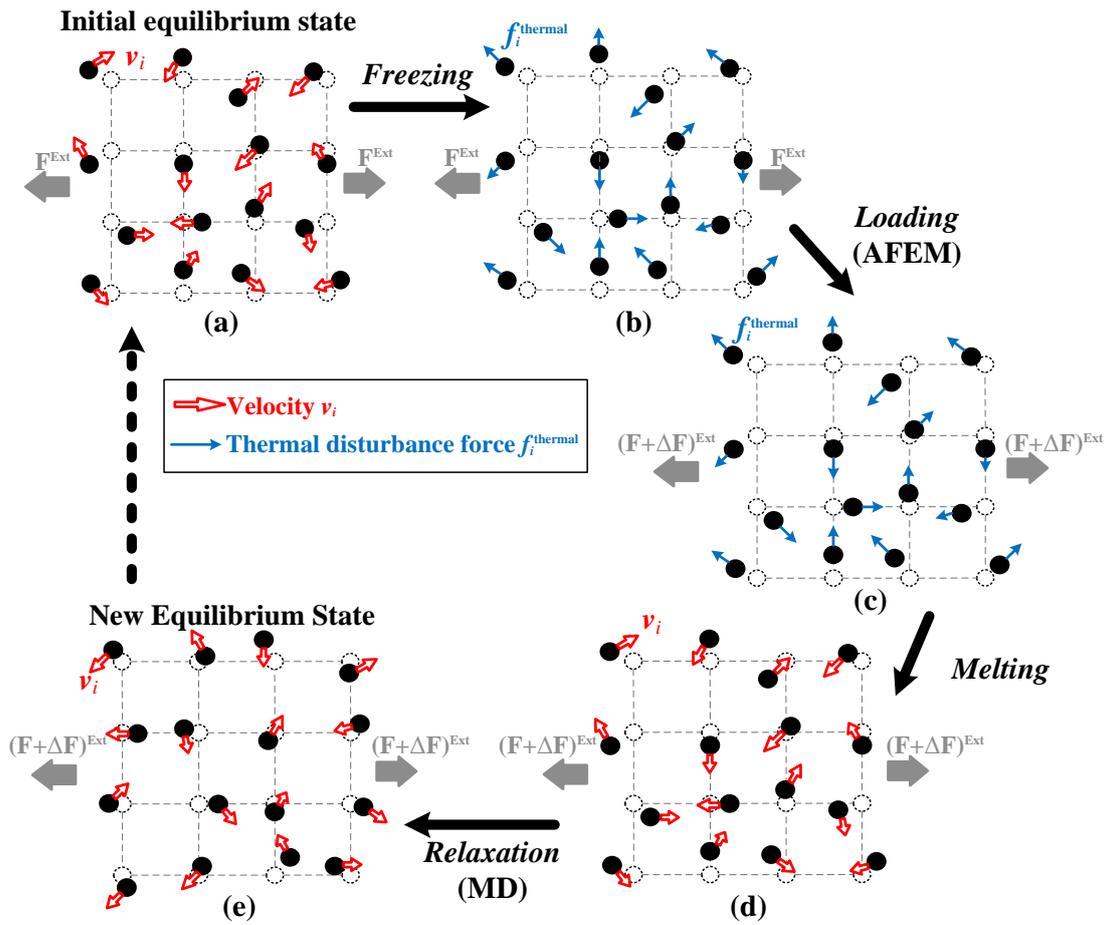

Figure 1 Schematic flow chart of hybrid molecular dynamics/atomic-scale finite element method for quasi-static atomistic simulations at finite temperature.

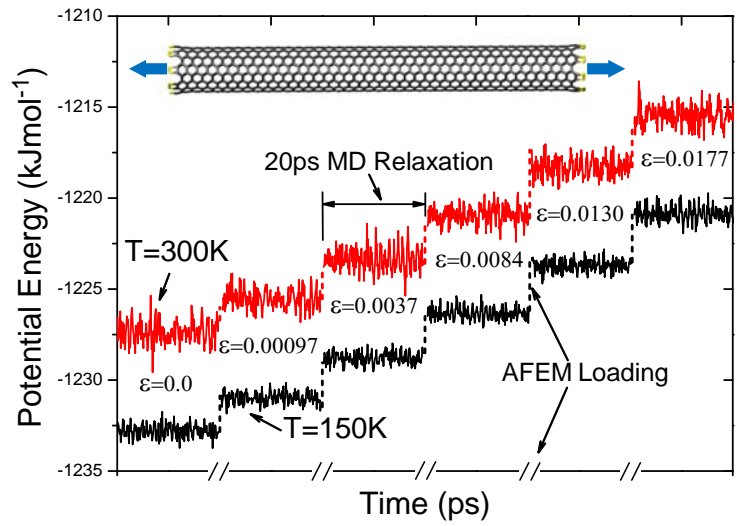

Figure 2 The potential energy variation of (8,8) CNT in step-by-step tensile test simulated by the hybrid MD/AFEM atomic quasi-static algorithm at $T$=150K and $T$=300K.

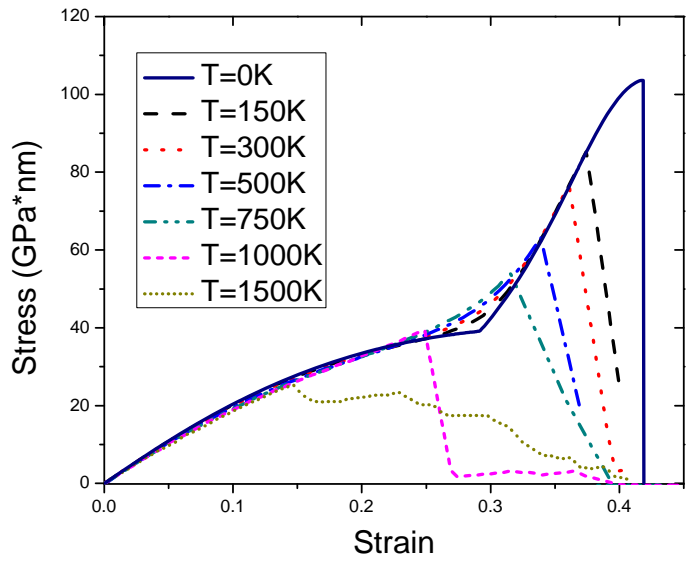

(a)

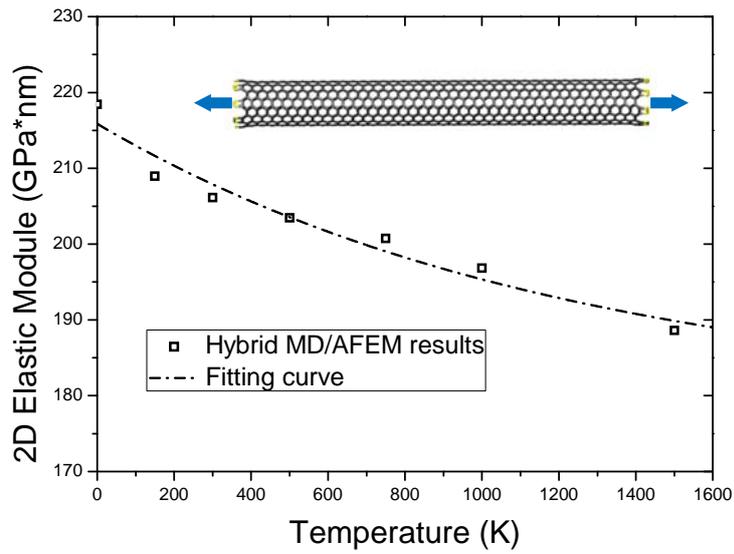

(b)

Figure 3 (a) The static 2D axial tensile stress as function of tensile strain for (8,8) CNT at different temperatures; (b) The temperature dependence of the tensile modulus of (8,8) single-wall CNT.

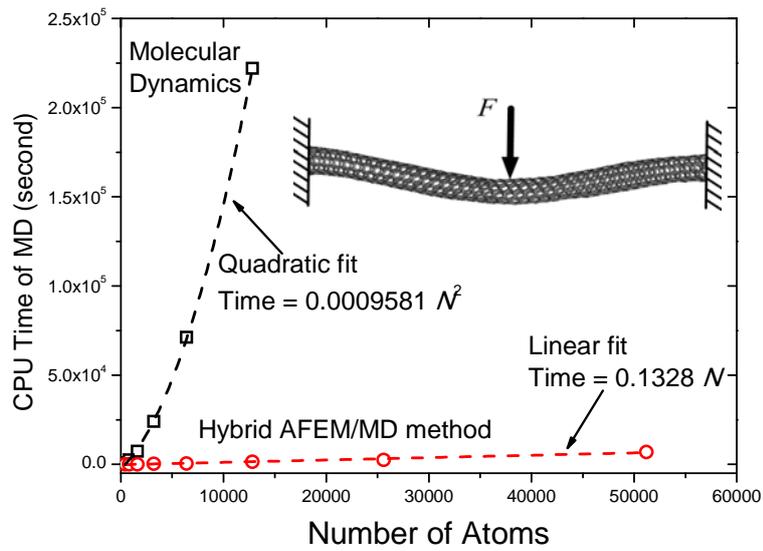

Figure 4   The required CPU time of the hybrid AFEM/MD method and the pure MD to acquire the steady deformed configurations of initially straight (5,5) armchair carbon nanotubes subject to the lateral force.

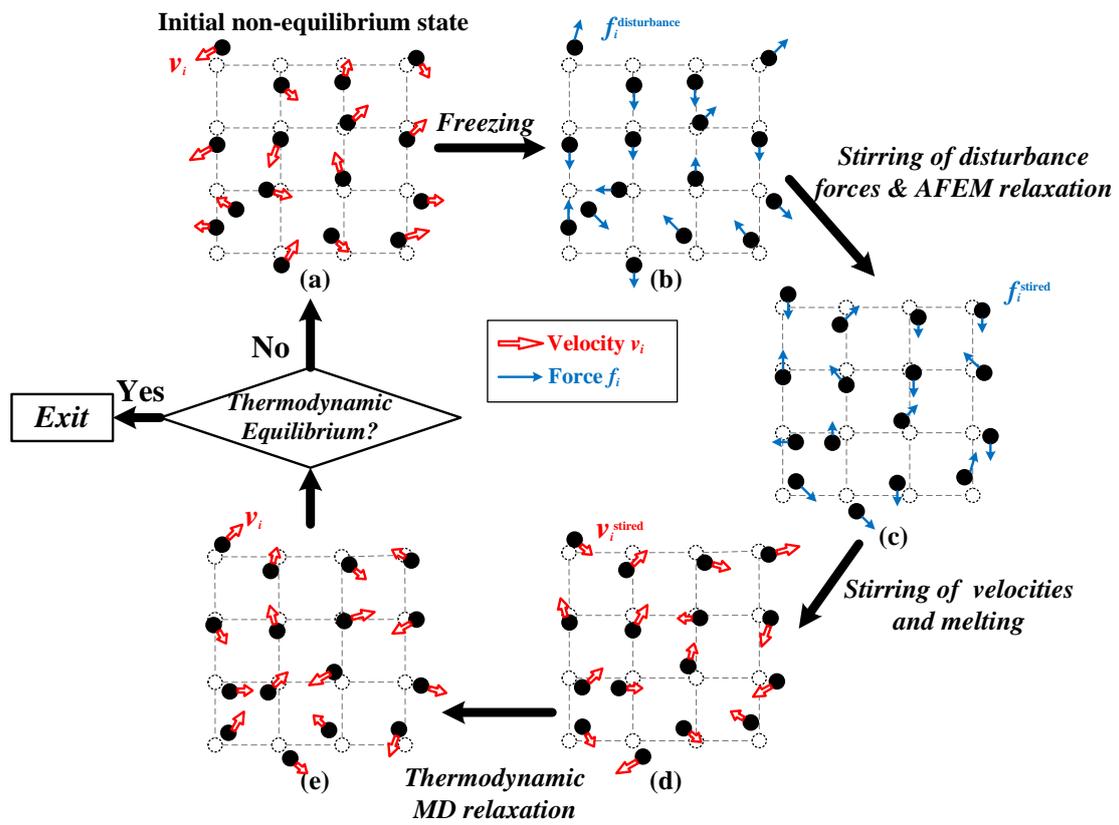

Figure 5 Schematic flow chart of stirring-accelerated MD/AFEM relaxation method for non-equilibrium molecular systems.

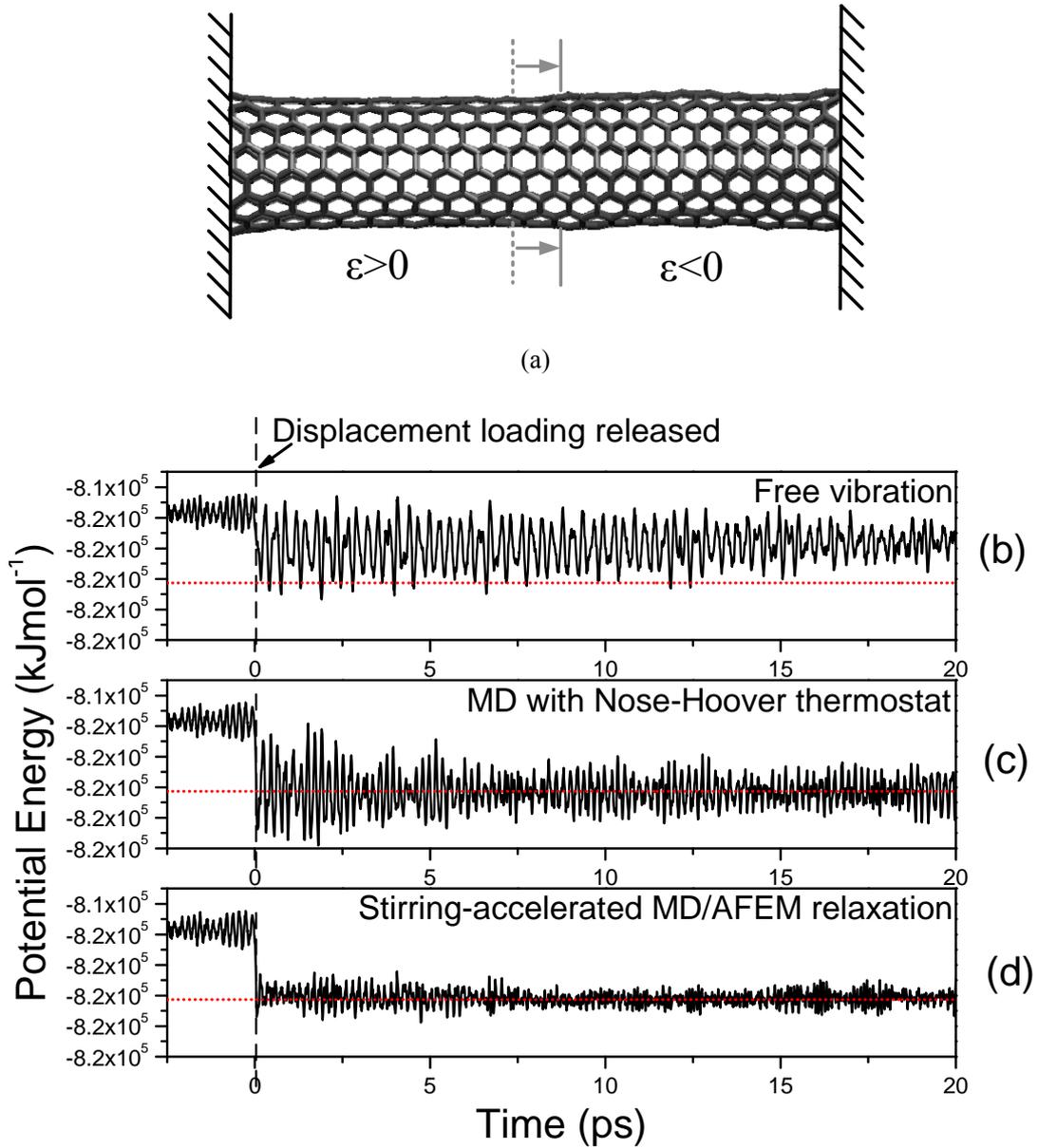

Figure 6 (a) Schematic diagram of an (8,8) single wall CNT subject to a axial displacement loading in the middle, and the displacement loading is suddenly released at 0ps. The variation of potential energy of CNT versus time of three relaxation simulations: MD without any interference (i.e., free vibration) (b), MD with Nose-Hoover thermostat at 300K (c), and stirring-accelerated MD/AFEM relaxation (d).